\documentclass[aps,prb,twocolumn,superscriptaddress,showpacs]{revtex4}
\usepackage{graphicx}% Include figure files
\usepackage{bm}
\usepackage{epsfig}
\usepackage{ulem}
\usepackage{color}
\usepackage{mathrsfs}
\usepackage{dcolumn}
\usepackage{setspace}
\usepackage{array}
\usepackage{amsmath}
\usepackage{amssymb}
\usepackage{gensymb}
\usepackage{bibentry,natbib}
\usepackage{booktabs}
\usepackage{url}
\usepackage{hyperref}
\begin{document}

\title{Doubled Shapiro steps in a topological Josephson junction}

\author{Yu-Hang Li}
\affiliation{International Center for Quantum Materials, School of Physics, Peking University, Beijing 100871,  China}

\author{Juntao Song}
\affiliation{Department of Physics and Hebei Advanced Thin Film Laboratory, Hebei Normal University, Shijiazhuang 050024, China}

\author{Jie Liu}
\email[]{jieliuphy@mail.xjtu.edu.cn}
\affiliation{Department of Applied Physics, School of Science, Xian Jiaotong University, Xian 710049, China}

\author{Hua Jiang}
\affiliation{College of Physics, Optoelectronics and Energy, Soochow University, Suzhou 215006, China}

\author{Qing-Feng Sun}
\affiliation{International Center for Quantum Materials, School of Physics, Peking University, Beijing 100871,  China}
\affiliation{Collaborative Innovation Center of Quantum Matter, Beijing 100871, China}
\affiliation{CAS Center for Excellence in Topological Quantum Computation, University of Chinese Academy of Sciences, Beijing 100190, China}

\author{X. C. Xie}
\affiliation{International Center for Quantum Materials, School of Physics, Peking University, Beijing 100871,  China}
\affiliation{Collaborative Innovation Center of Quantum Matter, Beijing 100871, China}
\affiliation{CAS Center for Excellence in Topological Quantum Computation, University of Chinese Academy of Sciences, Beijing 100190, China}

\date{\today}

\begin{abstract}

We study the transport properties of a superconductor-quantum spin Hall insulator-superconductor hybrid system in the presence of microwave radiation. Instead of adiabatic analysis or use of the resistively shunted junction model, we start from the microscopic Hamiltonian and calculate the d.c. current directly with the help of the non-equilibrium Green's function method. The numerical results show that (i) the $I$-$V$ curves of background current due to multiple Andreev reflections exhibit a different structure from those in the conventional junctions, and (ii) all Shapiro steps are visible and appear one by one at high frequencies, while at low frequencies, the steps evolve exactly as the Bessel functions and the odd steps are completely suppressed, implying a fractional Josephson effect.

\end{abstract}

\maketitle
\section{Introduction}
The Majorana bound state (MBS), which harbors non-Abelian statistics, has recently attracted extensive interest for its potential applications in fault-tolerant topological quantum computation\cite{Stern2010,Simon2008,KITAEV2003,Sarma2005,Freedman2005}. The realization of these states was first expected theoretically by Kitaev in a one-dimensional spinless $p$-wave superconducting chain\cite{Kiteav2001}. Unfortunately, despite that $p$-wave pairing is scarce in nature due to spin degeneracy, the inevitable `Majorana fermion doubling problem'\footnote{Suppose the spinless fermions are replaced by spinful ones. In this case each end of the Kiteav chain supports two Majorana zero modes, or equally, an ordinary fermionic zero mode, the energy of which will move away from zero due to some inevitable effects such as spin-orbital coupling.} made it impossible to realize in experiments\cite{Fidkowski2010,Fidkowski20111,Niu2012}. Soon afterward, many schemes for engineering Kitaev's ideal model in condensed material systems were put into practice\cite{Fidkowski20112,Sau2011,Fu2008,Lutchyn2010,Oreg2010,Cook2011}.
A conceptual breakthrough came in 2009 when Fu and Kane proved that topological junctions between superconductors mediated by a quantum spin Hall insulator (QSHI) can stabilize those MBSs at their interfaces\cite{Fu2009}.
In this system, effective $p$-wave pairing can be achieved by superconducting proximity effects combined with time reversal symmetry breaking.  
Furthermore, the `Majorana fermion doubling problem' is automatically circumvented because there exists only one pair of Fermi points as long as the Fermi level does not intersect the bulk bands.
In addition, MBSs are also proposed to exist in other systems, for example, as quasiparticle excitations of the quantum Hall state at filling factor $\nu=5/2$\cite{Moore1991,Read2000}, in the vortices of the intrinsic $p$-wave superconductor $Sr_2RuO_4$\cite{Sarma2006}, and in cold atom systems\cite{Gurarie2005,Tewari2007}.

Experimental probes of these MBSs can be achieved by measurement of the fractional Josephson effect\cite{Alicea2012,Beenakker2013,Yang2016}. The coupling of two MBSs $\gamma_{1}$ and $\gamma_{2}$ localized at the interfaces of a topological Josephson junction allows the tunneling of half Fermion pairs, and in turn yields a $4\pi$ periodic supercurrent $I_{4\pi}sin\left(\phi/2\right)$, namely the fractional Josephson effect. As a result, in the presence of a d.c. bias voltage $V_{0}$, one would expect an a.c. Josephson current at half Josephson frequency $\omega_{0}/2=eV_{0}/\hbar$ accompanied by radio-frequency radiation of the same frequency\cite{Lee2014,Badiane2011,Jose2012,Deacon2017}.
Moreover, supplementing the junction with an rf emission of frequency $\omega$, a current measurement will find plateaus of the voltage steps, also known as Shapiro steps, emerging only when  $2eV_{0}/\hbar=2n\omega$\cite{Badiane2011,Dom2012,Dom2017,Sau2017,Jose2012}, where $n$ is an integer, leading to an even-odd effect with all odd steps disappeared.
A second type of fractional Josephson effect, which exhibits $4\pi$ periodicity in both superconducting phases of the left and right leads, may arise if the barrier material in the Josephson junction is also a superconductor\cite{Alicea2012,Liang2011}.
Some recent experiments were performed to explore this even-odd effect in superconductor-quantum spin Hall insulator-superconductor S-QSHI-S Josephson junctions\cite{Wiedenmann2015,Bocquillon2016} and several other systems\cite{Rokhinson2012,Li2017} which are believed may hold MBSs. Interestingly, the results show a strong frequency dependence.
Thus far, only the resistively shunted junction model has been considered to understand this effect\cite{prb95195430}. Under this approach, the system is simplified as a circuit with a Josephson junction shunted by a resistance $R$, which can be described by an equation of motion: $I_{0}+I_{ac}sin\left(\omega t\right)=V/R+I\left(\phi\right)$. The $4\pi$ periodic term $I\left(\phi\right)=I_{4\pi}sin\left(\phi/2\right)$ in this equation phenomenologically leads to an even-odd effect in Shapiro steps. However, a microscopic mechanism of the direct connection between the presence of MBSs and this even-odd effect is lacking, and the underlying physics of this effect's been exhibiting only at low frequencies in experiments need to be understood.

In this paper, we study the transport properties of an S-QSHI-S Josephson junction.  Using the non-equilibrium Green's functions method, we calculate the tunneling current based on a tight-binding Hamiltonian. Our numerical results show that the $I$-$V$ curves of the background currents exhibit interesting subharmonic gap structure, which is caused by the multiple Andreev reflections  (MARs). Different from the conventional Josephson junctions, the presence of MBSs reduces the gap from $2\Delta$ to $\Delta$, and therefore, the I-V curves have singularities at voltages $eV_{0}=\left(\Delta\mp k\hbar\omega\right)/n$ rather than $eV_{0}=\left(2\Delta\mp k\hbar\omega\right)/n$, with $n$, $k$ being integer numbers. On the other hand, we find that the Shapiro steps appear one by one and have complicated oscillation patterns at higher frequencies due to the nonadiabatic process. However, at low frequencies, the steps evolve exactly as the Bessel functions but with the odd steps suppressed strongly, in agreement with the recent experimental results in Ref.~[\onlinecite{Bocquillon2016}].

The rest of this paper is organized as follows. In Sec.~\ref{mf}, we introduce our model Hamiltonian and deduce the equation of the supercurrent by virtue of the non-equilibrium Green's functions method. In Sec.~\ref{results}, we focus our results on the d.c. current and study the back-ground current and the Shapiro steps in detail. Finally, a brief summary is presented in Sec.~\ref{summary}. 

\section{Model and Formalism}
\label{mf}
\begin{figure}[htbp]
  \centering
  \includegraphics[width=0.45\textwidth]{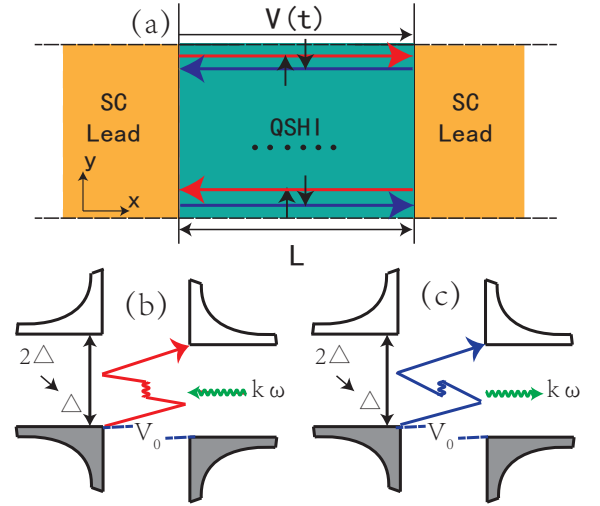}
  \caption{(color online).
             (a) Schematic for the S-QSHI-S device with voltage $V\left(t\right)=V_{0}+V_{1}sin\left(\omega t\right)$.
             (b, c) Schematic diagrams for three-order MAR mediated by absorbing or emitting k photons, respectively.}
\label{fig:intro}
\end{figure}

We consider a voltage biased S-QSHI-S Josephson junction in the presence of microwave radiation as shown in Fig.~\ref{fig:intro}(a).
To proceed, this external field is simulated with a time-dependent voltage
$V\left(t\right)=V_{0}+V_{1}sin\left(\omega t\right)$.
Then after a unitary transformation\cite{Sun2000}, the system can be described by the following Hamiltonian:
\begin{equation}
\mathcal{H}=\sum_{\alpha=L,R}\mathcal{H}_{\alpha}\left(t\right)+\mathcal{H}_{C}+\mathcal{H}_{T}.
\label{eq:th}
\end{equation}
Here, $\mathcal{H}_{\alpha}\left(\alpha=L,R\right)=\sum_{k,\sigma}\epsilon_{{\alpha}k}a_{{\alpha}k\sigma}^{\dagger}a_{{\alpha}k\sigma}+\Delta\left(a_{{\alpha}k\downarrow}a_{{\alpha}-k\uparrow}+\rm{H.c.} \right)$ are the BCS Hamiltonians of both the left and the right $s$-wave superconducting leads, where $a_{{\alpha}k{\sigma}}^{\dagger}\left(a_{{\alpha}k{\sigma}}\right)$ are the creation (annihilation) operators of electrons in the $\alpha$ lead with momentum $k$ and spin $\sigma$, $\epsilon_{{\alpha}k}$ is the kinetic energy, and $\Delta$ is the common superconducting energy gap shared by both leads.
As the transport properties are dominated by the Helical edge states of the central QSHI\cite{Song2016,Hart2014}, this part can be described by an effective one-dimensional Hamiltonian\cite{Zhou2017}, which in the Nambu representation is $\mathcal{H}_{C}=\frac{-iV_{F}}{2a}\bold{\Psi}_{i}^{\dagger}\bold{\sigma}_{z}\bold{\tau}_{z}\bold{\Psi}_{i+\delta x}-\mu\bold{\Psi}_{i}^{\dagger}\bold{\tau}_{z}\bold{\Psi}_{i}+M\bold{\Psi}_{i}^{\dagger}\bold{\sigma}_{x}\bold{\Psi}_{i}$.  $a$ is the lattice constant, $\bold{\Psi}_{i}^{T}=\left[\psi_{i\uparrow},\psi_{i\downarrow},
\psi_{i\downarrow}^{\dagger},-\psi_{i\uparrow}^{\dagger}\right]$ are the edge states, $V_{F}$ and $\mu$ are the velocity and the chemical potential of edge states, $M$ is the Zeeman erengy caused by an external magnetic field, and $\bold{\sigma}_{j}$ and $\bold{\tau}_{j}$ are the Pauli matrices acting in the spin and Nambu spaces, respectively.
The last term in Eq.~(\ref{eq:th}), representing the time-dependent coupling between the superconducting leads and the central part, has the form $\mathcal{H}_{T}=\sum_{\alpha}\left(a_{\alpha k\uparrow}^{\dagger},a_{\alpha k \downarrow}\right)\bold{t}_{\alpha C}\bold{\Psi}_{i}+\rm{H.c.}$. The coupling matrix $\bold{t}_{\alpha C}\left(t\right)=\begin{pmatrix} \tilde{t}_{\alpha} & \tilde{t}_{\alpha} & 0 & 0 \\ 0&0& -\tilde{t}_{\alpha}^{*} & -\tilde{t}_{\alpha}^{*} \end{pmatrix}$, where $\tilde{t}_{\alpha}=t_{c}exp\left\{i\left[{\phi_{\alpha}}/{2}+z_{\alpha}t+p_{\alpha} cos\left(\omega t\right)\right]\right\}$, with $t_{c}$ the coupling strength, $\phi_{L,R}$ the initial superconducting phases of the left and right leads, $z_{\alpha}=eV_{0}^{\alpha}/\hbar$ the d.c. voltage, and $p_{\alpha}=eV_{1}^{\alpha}/\hbar\omega$ the radiation power.
In fact, by making a unitary transformation with $\tilde{a}^{\dagger}_{\alpha k\uparrow/\downarrow}
=\frac{\sqrt{2}}{2}(a^{\dagger}_{\alpha k\uparrow/\downarrow} \pm a^{\dagger}_{\alpha k\downarrow/\uparrow})$, $\mathcal{H}_{T}$ can reduce to
$\mathcal{H}_{T} = \sqrt{2} \tilde{t}_{\alpha} \tilde{a}^{\dagger}_{\alpha k\sigma}\psi_{i\sigma} +\rm{H.c.}$, the same as for the normal barriers.

The total current from the left superconducting lead can be calculated from the evolution of the electron number operator ${N}_{L}=\sum_{k\sigma}a_{Lk\sigma}^{\dagger}a_{Lk\sigma}$ in that lead
\begin{equation}
\begin{split}
I_{L}\left(t\right)&=-e{\langle}\frac{dN_{L}}{dt}{\rangle}=\frac{ie}{\hbar}\left\langle{\left[\sum_{k,\sigma}a_{L{k}\sigma}^{\dagger}a_{L{k}\sigma}, \mathcal{H}\right]}\right\rangle\\
&=\frac{e}{\hbar}\rm{Tr}\left[\bold{\Gamma}_{z}\bold{G}_{CL}^{<}\left(t,t\right)\bold{t}_{LC}+\rm{H.c.}\right],
\end{split}
\label{eq:current}
\end{equation}
where $\bold{\Gamma}_{z}=\bold{\sigma}_{z}\bigotimes \bold{I}_{2}$, with $\bold{I}_{2}$ the $2\times2$ identity matrix, and $\bold{G}_{CL}^{<}\left(t,t\right)$ is the distribution Green's function, which satisfies the relation: $\bold{G}_{CL}^{<}\left(t,t\right)=\int dt_{1}\left[\bold{G}^{r}\left(t,t_{1}\right)\bold{t}_{LC}^{\dagger}\bold{g}_{L}^{<}\left(t_{1},t\right)+\bold{G}^{<}\left(t,t_{1}\right)\bold{t}_{LC}^{\dagger}\bold{g}_{L}^{a}\left(t_{1},t\right)\right]$.  $\bold{g}_{L}^{<,a}\left(t_{1},t_{2}\right)$ are the surface Green's functions of the uncoupled superconducting lead, and $\bold{G}^{r,<}\left(t_{1},t_{2}\right)$ are the retarded and distribution Green's functions of the central QSHI part.
For convenience,
we take the left superconducting lead as the potential ground. Thus the current can be rewritten as
\begin{equation}
\begin{split}
I_{L}\left(t\right)=&\frac{2e}{\hbar}\rm{Im}\int_{-\infty}^{t} dt_{1}\int \frac{d\epsilon}{2\pi} e^{i\epsilon \left(t-t_{1}\right)}\rm{Tr}\left\{\bold{\Gamma}_{z}\left[\bold{G}^{r}\left(t,t_{1}\right)\right.\right.\\
&\left.\left.\times\bold{\Sigma}_{L}^{<}\left(\epsilon\right)+\bold{G}^{<}\left(t,t_{1}\right)\bold{\Sigma}_{L}^{a}\left(\epsilon\right)\right]\right\},
\end{split}
\end{equation}
where $\bold{\Sigma}_{L}^{<,a}\left(\epsilon\right)=\bold{t}_{LC}^{\dagger}\bold{g}_{L}^{<,a}\left(\epsilon\right)\bold{t}_{LC}$ are the distribution and retarded self-energies  due to coupling to the left superconducting lead. The exact retarded Green's functions of the uncoupled superconducting lead read $\bold{g}_{L,R}^{r}\left(\epsilon\right)=2\pi\rho\beta\left(\epsilon\right)\left[\bold{I}_{2}+\Delta/\epsilon \bold{\sigma}_{x}\right]$, where the corresponding BCS density of states $\beta\left(\epsilon\right)$ is defined as:  $\beta\left(\epsilon\right)=\epsilon/\left(i\sqrt{\Delta^2-\epsilon^{2}}\right)$ for $\Delta>\left|\epsilon\right|$, and $\beta\left(\epsilon\right)=\left|\epsilon\right|/\sqrt{\epsilon^{2}-\Delta^2}$ for $\Delta<\left|\epsilon\right|$. In addition, the normal density of states $\rho$ is assumed to be independent of the energy $\epsilon$. The advanced Green's functions $\bold{g}_{L,R}^{a}$ are the complex conjugates of the retarded Green's function and $\bold{g}_{L,R}^{<}=f\left(\epsilon\right)\left(\bold{g}_{L,R}^{a}-\bold{g}_{L,R}^{r}\right)$, where $f\left(\epsilon\right)=1/\left(1+e^{\epsilon/k_{B}T}\right)$ is the Fermi distribution function.

In order to obtain the Green's function, following the method in Ref.~[\onlinecite{Cuevas2002}], we perform a Fourier transform with respect to the temporal arguments, $\bold{G}\left(t_{1},t_{2}\right)=1/2\pi\int d\epsilon_{1}\int d\epsilon_{2}e^{-i\epsilon_{1}t_{1}}e^{i\epsilon_{2}t_{2}}\bold{G}\left(\epsilon_{1},\epsilon_{2}\right)$.  Because the phase difference of the junction is a time dependent periodic function with two periods $T_1=2\pi/\omega_0$ and $T_2=2\pi/\omega$, where $\omega_{0}=2e\left|V_0\right|$, $\bold{G}\left(\epsilon_{1},\epsilon_{2}\right)$ satisfies the following relation:$\bold{G}\left(\epsilon_{1},\epsilon_{2}\right)=\sum_{m,n}\bold{G}\left(\epsilon_{1},\epsilon_{1}+m\omega_{0}+n\omega\right)\delta\left(\epsilon_{2}-\epsilon_{1}-m\omega_{0}-n\omega\right)$, where $m$, $n$ are integer numbers. To simplify the mathematical expression of the supercurrent, we introduce the quantities $\bold{G}_{mn}^{kl}\equiv \bold{G}\left(\epsilon+m\omega_{0}+k\hbar\omega,\epsilon+n\omega_{0}+l\hbar\omega\right)$.  Finally, the current can now be expanded as $I\left(t\right)=\sum_{n,m}I_{n}^{m}exp\left[i\left(n\omega_{0}t+m\omega t\right)\right]$, where the current amplitudes $I_{n}^{m}$ can be expressed as:

\begin{equation}
\begin{split}
I_{n}^{m}=\frac{2e}{\hbar}\rm{Im}\int &\frac{d\epsilon}{2\pi}\rm{Tr}\left\{\bold{\Gamma}_{z}\left[\bold{G}_{-n0}^{r;-m0}\left(\epsilon\right)\bold{\Sigma}_{L}^{<}\left(\epsilon\right)\right.\right.\\
&\left.\left.+\bold{G}_{-n0}^{<;-m0}\left(\epsilon\right)\bold{\Sigma}_{L}^{a}\left(\epsilon\right)\right]\right\}.
\end{split}
\end{equation}
At this point, the calculation of the supercurrent has been reduced to the calculation of the Fourier components of the retarded and distribution Green's functions, which can be determined by numerically solving the Dyson equation and the Keldysh equation
\begin{subequations}
\begin{equation}
\bold{G}_{mn}^{r;kl}=\bold{g}_{m}^{r;k}\delta_{mn}\delta_{kl}+\sum_{i,j}\bold{G}_{mi}^{r;kj}\bold{\Sigma}_{in}^{r;jl}{g}_{n}^{r;l},
\label{Dyson}
\end{equation}
\begin{equation}
\bold{G}_{mn}^{<;kl}=\sum_{i_{1},i_{2},j_{1},j_{2}}\bold{G}_{mi_{1}}^{r;kj_{1}}\bold{\Sigma}_{i_{1}i_{2}}^{<;j_{1}j_{2}}\bold{G}_{i_{2}n}^{a;j_{2}l},
\end{equation}
\end{subequations}
where $\bold{g}_{m}^{r;k}=\bold{g}^{r}\left(\epsilon+m\omega_{0}+k\omega\right)$, and $\bold{\Sigma}_{mn}^{r,<;kl}=\bold{\Sigma}_{L;mn}^{r,<;kl}+\bold{\Sigma}_{R;mn}^{r,<;kl}$. The Fourier components of the self energies  $\bold{\Sigma}_{L,R;mn}^{r,<;kl}$ adopt the forms
\begin{widetext}
\begin{subequations}
\begin{equation}
\bold{\Sigma}_{L;nm}^{r;kl}=-{i}\pi t_{c}^2\beta_{n}^{k}{\bold{\Sigma}}_{L}\left(\epsilon_{n}^{k}\right)\delta_{nm}\delta_{kl},
\end{equation}
\begin{equation}
\bold{\Sigma}_{R;nm}^{r;kl}=-i\pi t_{c}^2\sum_{j}J_{k-j}\left(p\right)J_{j-l}\left(p\right)\begin{pmatrix}\left(-1\right)^{j}i^{-k-l}\beta_{n+1/2}^{j}\delta_{nm}&-i^{k-l}\beta_{n+1/2}^{j}\frac{\Delta}{\epsilon_{n+1/2}^{j}}e^{-i\phi_{R}}\delta_{n,m-1}\\i^{l-k}\beta_{n-1/2}^{j}\frac{\Delta}{\epsilon_{n-1/2}^{j}}e^{i\phi_{R}}\delta_{n,m+1}&\left(-1\right)^{j}i^{k+l}\beta_{n-1/2}^{j}\delta_{nm}\end{pmatrix}\bigotimes\bold{M},
\end{equation}
\end{subequations}
\begin{subequations}
\begin{equation}
\bold{\Sigma}_{L;nm}^{<;kl}=2i\pi t_{c}^2\gamma_{n}^{k}{\bold{\Sigma}}_{L}\left(\epsilon_{n}^{k}\right)\delta_{nm}\delta_{kl},
\end{equation}
\begin{equation}
\bold{\Sigma}_{R;nm}^{<;kl}=2i\pi t_{c}^2\sum_{j}J_{k-j}\left(p\right)J_{j-l}\left(p\right)\begin{pmatrix}\left(-1\right)^{j}i^{-k-l}\gamma_{n+1/2}^{j}\delta_{nm}&-i^{k-l}\gamma_{n+1/2}^{j}\frac{\Delta}{\epsilon_{n+1/2}^{j}}e^{-i\phi_{R}}\delta_{n,m-1}\\i^{l-k}\gamma_{n-1/2}^{j}\frac{\Delta}{\epsilon_{n-1/2}^{j}}e^{i\phi_{R}}\delta_{n,m+1}&\left(-1\right)^{j}i^{k+l}\gamma_{n-1/2}^{j}\delta_{nm}\end{pmatrix}\bigotimes\bold{M},
\end{equation}
\end{subequations}
\end{widetext}
where $\epsilon_{n}^{k}=\epsilon+n\omega_{0}+k\omega$, $\beta_{i}^{j}=\beta\left(\epsilon+i\omega_{0}+j\omega\right)$, $\gamma_{i}^{j}=f_{i}^{j}\beta_{i}^{j}$,   $\bold{M}=\bold{\sigma}_{0}+\bold{\sigma}_{x}$, and $J_{n}\left(p\right)$ is the first kind of Bessel function of order $n$, with $p=p_{R}-p_{L}$ denoting the radiation power. Finally, the time-dependent supercurrent  can be calculated without further complications.

\section{Results and discussions}
\label{results}
In this paper, we focus on the d.c. current, which consists of two parts: the background current $I_{0}^{0}$ and the Shapiro steps $I_{n}^{m}$. Notice that the Shapiro steps depend on the average value of the initial phase difference $\phi_{0}=\phi_{R}-\phi_{L}$. In the following, we take the superconducting energy gap $\Delta$ as the energy unit. The parameters of the central Hamiltonian are $a=5$nm, $L=100$nm, $\hbar V_{F}=200\Delta$, $M=1.5\Delta$ and $\mu=0$ so all energies are measured from the chemical potential. The coupling strength between the central QSHI part and the superconducting leads takes $t_{c}=1.8\Delta$ and the temperature is set as zero in our detailed calculations.
Here, the transmission probability is defined as $D=1/\left[1+sinh^2\left(ML\right)\right]${\cite{Fu2009}}. We fix the system parameters unless otherwise specified.

\begin{figure}[htbp]
  \centering
  \includegraphics[width=0.46\textwidth]{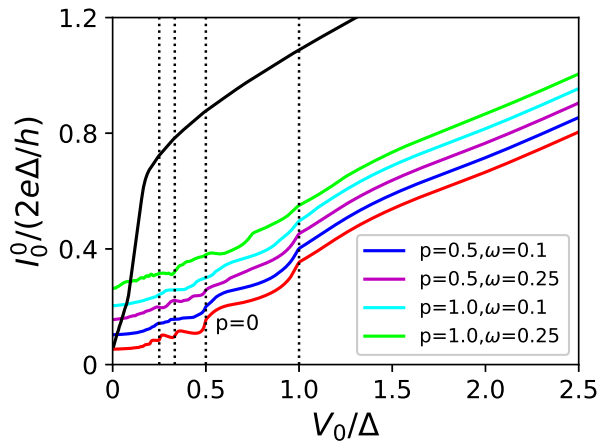}
  \caption{(color online). Background current $I_{0}^{0}$ as a function of the DC voltage $V_{0}$ with different parameters. Colorful curves are displayed to make the figure clear.
  Red curve, the case $p=0$; black curve. $M=0,p=0$. Here, $\phi_{L}=\phi_{R}=0$.}
\label{fig:back}
\end{figure}

Let us start by analyzing the background current. The key feature of the background current is that its $I$-$V$ curves have some singularities at discrete voltages $eV_{0}=\left(2\Delta\mp k\hbar\omega\right)/n$, which originates from an $n$-order MAR mediated by absorbing [Fig.~\ref{fig:intro}.(b)] or emitting [Fig.~\ref{fig:intro}.(c)] $k$ photons with their probabilities proportional to $J_{k}\left(p\right)$\cite{Cuevas2002}. After this photo-assisted MAR process, a quasi-particle acquires the energy $neV\pm k\hbar\omega$, and singularities appear simultaneously when this energy can overcome the energy gap $2\Delta$ between the occupied and the empty states. However, in topological Josephson junctions, because of the presence of the MBSs, the energy gap reduces to $\Delta$, and thus singularities should appear at $eV_{0}=\left(\Delta\mp k\hbar\omega\right)/n$ instead\cite{Badiane2011,P13}.

In Fig.~\ref{fig:back}, we plot the $I-V$ curves of the background current with different radiation power $p$ and frequency $\omega$. It can be clearly seen that when $p=0$ (red line), which means in the absence of microwave radiations, the curve exhibits gap structures at voltages $eV_{0}=\Delta/n$, while with microwave radiation added, the curves show rich subgap structures with more singularities appearing at $eV=\left(\Delta\mp k\hbar\omega\right)/n$(see the sign $eV_{0}=0.75$ of the lime-green line with $p=1.0$, $\omega=0.25$). This distinct structure strongly indicates the presence of MBSs.
Note that the black curve in Fig.~\ref{fig:back} corresponds to the case of $M=0$ and $p=0$,
in which the junction is like a conventional one due to the lack of localized MBSs. Naturally, because of the conducting helical edge, this junction is totally transparent with the transmission probability $D=1$, leading to a sharply increasing $I$-$V$ curve as plotted in the figure\cite{Cuevas1996}.

\begin{figure}[htbp]
  \centering
  \includegraphics[width=0.48\textwidth]{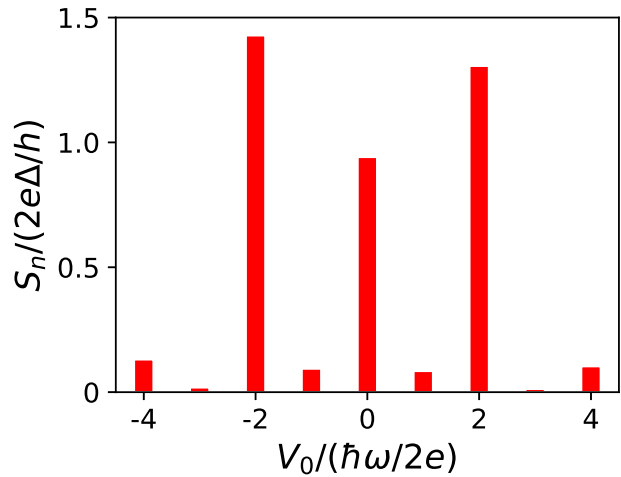}
  \caption{(color online). Bar plots of the heights of the Shapiro steps $S_n$ from $n=-4$ to $n=4$ at frequency $\omega=0.02\Delta$. Here, the radiation power is $p=\Delta$.}
\label{fig:bar}
\end{figure}

Now we move on to the Shapiro steps. These steps, arising from the phase locking between the harmonics of the a.c. Josephson frequency $\omega_{0}$ and the microwave radiation frequency $\omega$,
have been reported extensively in conventional Josephson junctions. Within an adiabatic approximation\cite{kopnin2009},   these steps can be understood as a consequence of  the nonsinusoidal current-phase relation. As stated above, the phase difference across the junction is $\phi\left(t\right)=\phi_{0}+zt+p\sin\left(\omega t\right)$. Substituting this into the Josephson's first equation and using the standard mathematical expansion of a sine in terms of the Bessel functions, one would expect the Shapiro steps to evolve exactly as $J_{n}\left(2p\right)$ and to appear at $z=n\omega$, where $n$ is an integer. In principle, the situations are different between conventional and topological Josephson junctions. In a conventional one, because the only carriers that are permitted to transmit through the central insulator part are Cooper pairs, Shapiro steps appear at $V_{0}=n\hbar\omega/2e$ with $z=2eV_{0}$. However, in a topological one, two MBSs, $\gamma_1$ and $\gamma_2$ are localized separately at the interfaces of the superconducting leads and the central QSHI region\cite{Fu2009}. Their strong coupling forms a $4\pi$ fractional Josephson effect and allows the tunneling of single electrons, therefore Shapiro steps can appear at double the voltage of the former one and exhibit an even-odd effect when $z=eV_0$ as plotted in Fig.~\ref{fig:bar}. Here, the heights of the Shapiro steps are defined as $S_{n}\equiv\left|I_{n}^{1}\right|$. It should be pointed out that the fractional Shapiro steps $I_{n}^{m}$ with $m>1$ are so small in our calculations that we have omitted them from the figure. Interestingly, the absence of the odd steps  here is not calculated by simply adding a $4\pi$ supercurrent in the RSJ model but the direct result of the non-equilibrium Green's function method based on the intrinsic Hamiltonian. Besides, this fundamental method can help us to further understand the frequency dependence of the even-odd effect.

\begin{figure}[htbp]
  \centering
  \includegraphics[width=0.48\textwidth]{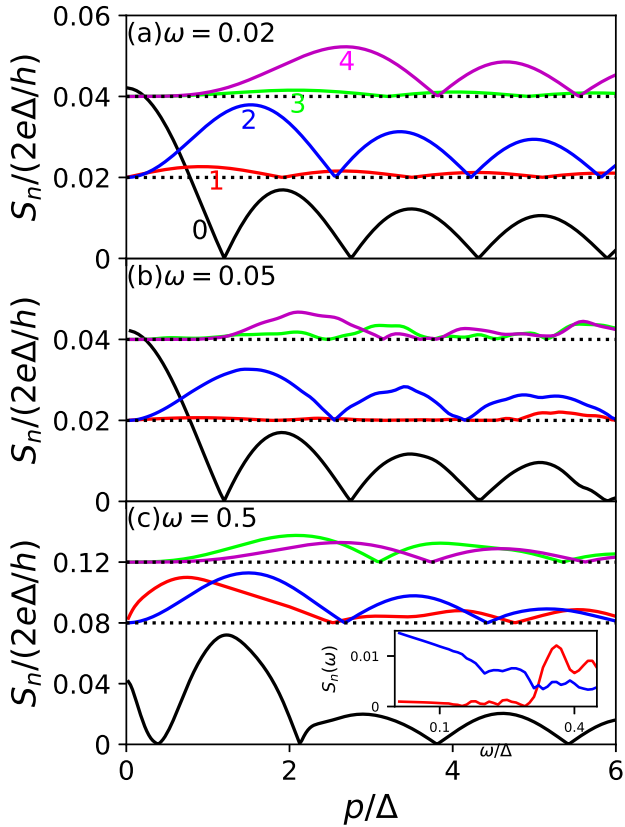}
  \caption{(color online). Heights of the first five Shapiro steps $S_{n}$ versus the radiation power $p$ at threefrequencies:  (a) $\omega=0.02$, (b) $\omega=0.05$, and (c) $\omega=0.5$. Here, some steps are vertical shifted for clarity. Inset:$S_1$ (red line) and $S_2$ (blue line) as the increase in the frequency with radiation power $p=\Delta$.}
\label{fig:shapiro}
\end{figure}

In Fig.~\ref{fig:shapiro}, we display the heights of the first five Shapiro steps $S_{n}$ as the increase in the radiation powers at three frequencies $\omega=0.02$, $\omega=0.05$ and $\omega=0.5$. As illustrated in Fig.~\ref{fig:shapiro}(a), the Shapiro steps coincide well with the Bessel functions except that the odd ones are strongly suppressed at frequency $\omega=0.02$, leading to an even-odd effect  as predicted and reported in some recent works\cite{Badiane2011,Jose2012,Wiedenmann2015,Bocquillon2016}. With the frequency increased to $\omega=0.05$, Fig.~\ref{fig:shapiro}(b) shows that although the first step is still heavily suppressed,  the higher order steps begin to appear. As a comparison, we plot the heights $S_{n}$ at a high frequency $\omega=0.5$ in Fig.~\ref{fig:shapiro}(c). Though the shapes deviate seriously from Bessel functions, all steps are visible and appear one by one as the increase in the radiation power. In general, the deviation results from a nonadiabatic process at high frequencies. The inset shows the heights $S_1$ and $S_2$ as the increase in the frequency $\omega$. We can clearly see that $S_1$ is suppressed heavily when $\omega<0.3\Delta$, which indicates that the even-odd effect can only be seen at low frequencies. {The reason is that the Andreev bound state may couple the continuum after absorbing a large frequency radiation, restoring a $2\pi$ periodicity.} For superconducting leads made of Al electrodes, our results are in general agreement with the experiment data in Ref.~[\onlinecite{Bocquillon2016}] if the time reversal symmetry is implicitly broken since the frequencies here are about $\omega=0.5$GHz [Fig.~\ref{fig:shapiro}(a)], $\omega=1.2$GHz [Fig.~\ref{fig:shapiro}(b)] and $\omega=12$GHz [Fig.~\ref{fig:shapiro}(c)]. The exact mechanism for the seemly perfect transmission in that work remains to be understood.

\section{summary}
\label{summary}
To summarize, we have studied the transport properties of an S-QSHI-S Josephson junction in the presence
of microwave radiation. Using non-equilibrium Green's functions, we calculate the d.c. supercurrent
at an arbitrary frequency starting from the initial tight-binding Hamiltonian.
The distinct singularities of the back-ground current prove that the presence of MBSs reduces the gap from $2\Delta$ to $\Delta$.
Furthermore, the even-odd effect of the Shapiro steps can only be seen at low frequencies.
Our theory provides a good explanation of the connection between the even-odd effect and the MBSs.

\section*{ACKNOWLEDGEMENTS}
This work was supported by the NBRP of China (Grant Nos. 2015CB921102 and 2014CB920901),
the National Key R and D Program of China (2017YFA0303301),
the NSF-China under Grant Nos. 11574007, 11204065, 11374219, 11574245, and 11534001,
and the Key Research Program of the Chinese Academy of Sciences (Grant No. XDPB08-4).

\begin{appendix}
\section{}
\label{seca}

In this Appendix, we present a numerical method for solving the Dyson equation [Eqn.~\ref{Dyson}] in Sec.~\ref{mf} above. 
In general, this equation can be solved literally. However, the process is very time-consuming. Inspired by the solution of the linear polynomial, we find that this equation can be solved similarly. Because the index in that equation can be any infinite large integers, cutoffs are necessary before we solve it. We assume that the lower index $-L\leq m,n\leq L$, and the upper index $-U\leq k,l\leq U$. Then the Dyson equation can be rewritten as
\begin{equation}
\bold{G}_{ij}^{r}=\bold{g}_{ij}^{r}\delta_{ij}+\sum_{i^{\prime}}\bold{G}_{ii^{\prime}}^{r}\bold{\Sigma}_{i^{\prime}j}^{r}\bold{g}_{jj}^{r},
\label{a1}
\end{equation} 
where $i=mU+k$, $j=nU+l$, and $i^{\prime}$ denotes the summation from $-LU$ to $LU$.
For a certain $i$, we replace the matrices above as $\bf{X}_{i^{\prime}}=\bf{G}^{r}_{ii^{\prime}}$, $\bf{A}_{i^{\prime}j}=\delta_{i^{\prime}j}-\bf{\Sigma}^r_{i^{\prime}j}\bf{g}^r_{jj}$, and $\bf{B}_{i^{\prime}j}=\bf{g}^{r}_{i^{\prime}j}\delta_{i^{\prime}j}$. Equation (.~\ref{a1}) can be equally written as
\begin{equation}
\hat{\bold{X}}^{T}\hat{\bold{A}}=\hat{\bold{B}},
\label{a2}
\end{equation}
where a hat denotes an array. Noting that every element in the array in Eq. (.~\ref{a2}) is also a matrix, $\hat{\bold{X}}$ can finally be obtained by block diagonalizing the array $\hat{\bold{A}}$. 

\begin{figure}[htbp]
  \centering
  \includegraphics[width=0.45\textwidth]{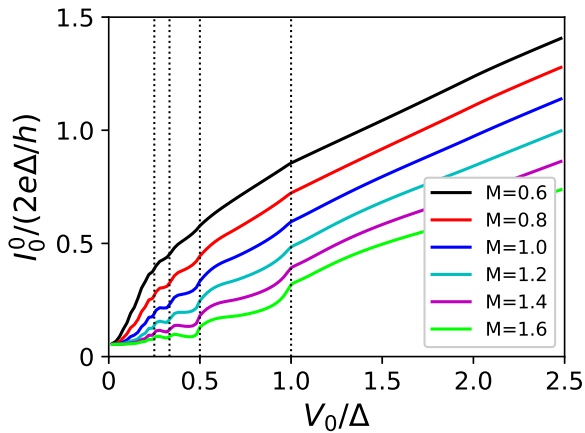}
  \caption{(color online).
             Background current $I_0^0$ as a function of the d.c. voltage $V_0$ at different Zeeman energies which correspond to different transmission probabilities without any radio-frequency radiation. Here, the other parameters are the same as in Fig. 2.}
\label{figs1}
\end{figure}

\section{}
\label{secb}
In this Appendix, we show the transmission probability dependence of the even-odd effect. As stated above, the transmission probability is defined as $D=1/\left[1+sinh^2\left(ML\right)\right]$, where $M$ is the Zeeman energy and $L$ is the length of the central QSHI region. Since we fix the system length as $L=100$nm, different Zeeman energies can be used to represent different transmission probabilities. In general, the fractional Josephson effect can be seen easily with a low transmission probability, or equally, a high Zeeman energy. In Fig.~\ref{figs1}, we display the back ground current $I^0_0$ versus the d.c. voltage $V_0$ at different Zeeman energies with $p=0$. All currents exhibit gap structures at voltages $eV_0=\Delta/n$, with $n$ an integer. Moreover, the structure can be seen more clearly as the increase of the Zeeman energy, which corresponds to the decrease in the transmission probability. This distinct structure strongly indicates that the superconducting gap is reduced from $2\Delta$ to $\Delta$, which is in very good agreement with the result in the main text. 

\begin{figure}[htbp]
  \centering
  \includegraphics[width=0.45\textwidth]{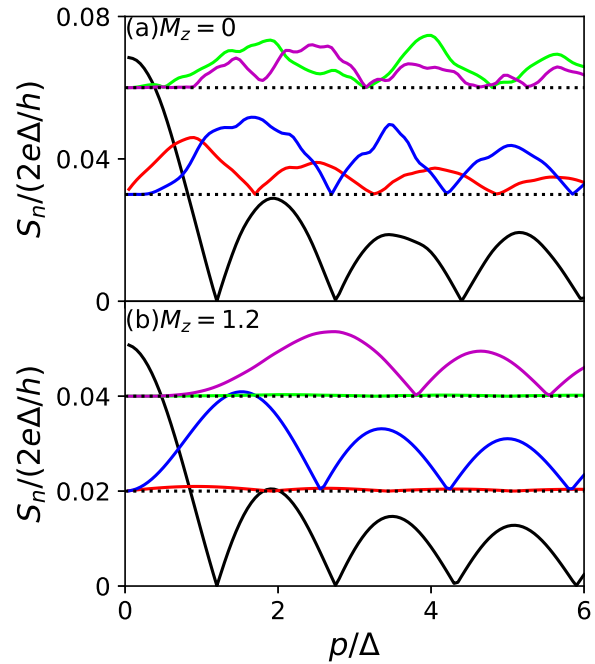}
  \caption{(color online).
             Heights of the first five Shapiro steps $S_n$ versus the radiation power $p$ at two transmission probabilities, (a) $M_z=0$ and (b) $M_z=1.2\Delta$. Here, the frequency is taken as $\omega=0.02$, and the other parameters are the same as in Fig.4.}
\label{figs2}
\end{figure}

In the experiment reported in [\onlinecite{Bocquillon2016}], the transmission probability seems to be $D=1$ since the time reversal symmetry is not explicitly broken. Theoretically, the supercurrent will restore a $2\pi$ periodicity due to a perfect transmission. Nevertheless, the experimental data contradict the existing theoretical proposals. In order to study this paradox, we show the heights of the first five Shapiro steps $S_n$ versus the increase in the radiation power $p$ at two transmission probabilities $M=0$ [Fig. 6(a) and $M=1.2\Delta$ [Fig. 6(b)]. As is clearly shown, the odd steps are only suppressed at $M_z=1.2\Delta$, which corresponds to a fractional transmission probability but different from that in the text, while at $M_z=0$, or equally $D=1$, all Shapiro steps are visible. This result generally agrees with the theoretical works but also contradicts the experiment data. The reason may be that the time reversal symmetry in the experiment is implicitly broken by some other effects such as puddles\cite{Deacon2017}. However, the exact mechanism still needs to be studied further.

\end{appendix}

\bibliography{citation}

\end{document}